\pdfoutput=1

\documentclass[11pt]{article}

\usepackage[final]{acl}

\usepackage{times}
\usepackage{latexsym}

\usepackage{microtype}
\usepackage{amsmath}
\usepackage{cleveref}
\usepackage{subfigure}
\usepackage{booktabs} 
\usepackage{multirow}

\usepackage{amsfonts}
\usepackage{adjustbox}
\usepackage{lineno}

\usepackage{amssymb}
\usepackage{mathtools}
\usepackage{amsthm}

\usepackage{cleveref}

\usepackage[T1]{fontenc}

\usepackage[utf8]{inputenc}

\usepackage{microtype}

\usepackage{inconsolata}

\usepackage{graphicx}

\title{Language-Codec: Bridging Discrete Codec Representations and Speech Language Models}

\author{
 \textbf{Shengpeng Ji\textsuperscript{1}\thanks{Equal contribution.}},
 \textbf{Minghui Fang\textsuperscript{1}\footnotemark[1]},
 \textbf{Jialong Zuo\textsuperscript{1}\footnotemark[1]},
 \textbf{Ziyue Jiang\textsuperscript{1}},
\textbf{Dingdong Wang} \\
 \textbf{Hanting Wang\textsuperscript{1}},
 \textbf{Hai Huang\textsuperscript{1}},
 \textbf{Zhou Zhao\textsuperscript{1}\thanks{Corresponding author.}}
\\
 \textsuperscript{1}~Zhejiang University ~~~~
}

\begin{document}
\maketitle
\begin{abstract}
In recent years, large language models have achieved significant success in generative tasks related to speech, audio, music, and other signal domains. A crucial element of these models is the discrete acoustic codecs, which serve as an intermediate representation replacing the mel-spectrogram. However, there exist several gaps between discrete \textbf{codecs} and downstream speech \textbf{language} models. Specifically, 1) Due to the reconstruction paradigm of the Codec model and the structure of residual vector quantization, the initial channel of the codebooks contains excessive information, making it challenging to directly generate acoustic tokens from weakly supervised signals such as text in downstream tasks. 2) numerous codebooks increases the burden on downstream speech language models. Consequently, leveraging the characteristics of speech language models, we propose \textbf{Language-Codec}. In the Language-Codec, we introduce a Masked Channel Residual Vector Quantization (MCRVQ) mechanism along with improved fourier transform structures and attention blocks, refined discriminator design to address the aforementioned gaps. We compare our method with competing audio compression algorithms and observe significant outperformance across extensive evaluations. Furthermore, we also validate the efficiency of the Language-Codec on downstream speech language models. Codes are available at
\url{https://github.com/jishengpeng/Languagecodec}.
\end{abstract}

\section{Introduction}
In recent times, significant achievements have been made by large-scale language models \cite{gpt3} in generative tasks involving such as multiple speaker speech syntheses \cite{valle,speartts,textrolspeech}, music  generation \cite{musiclm}, and audio generation \cite{audiogen}. This success can largely be attributed to the utilization of discrete acoustic codec representations produced by neural codec models \cite{soundstream,encodec}, which enable powerful transformer-based sequence-to-sequence modeling approaches for audio generation. The primary objective of discrete codec models is to convert a high-resolution audio signal (e.g., audio sampled at 44 kHz per second) into the two-dimensional discrete space. This transformation allows for the maximal compression of the speech signal in the time and frequency domains while maintaining excellent audio reconstruction quality.

Currently, most end-to-end discrete codec models \cite{soundstream,encodec,hificodec} typically adopt a three-stage structure consisting of an encoder, a Residual Vector Quantization (RVQ) module, and a decoder. The encoder performs downsampling of the audio signal in the time domain to obtain compressed audio frames. Each compressed audio frame is then quantized by a series of quantizers, with each quantizer operating on the residual of the previous one. The number of quantizers determines the overall bitrate. The decoder, on the other hand, performs upsampling in the time domain to reconstruct the audio signal from the quantizer outputs.

While most codec models strive to optimize their architecture \cite{kumar2023high,encodec,vocos}, resulting in satisfactory audio reconstruction quality,
there are still areas worth investigating in the construction of a discrete acoustic codec space that facilitates downstream speech language model modeling \cite{valle,soundstorm,speartts}.
Specifically, we believe that there exist gaps between discrete codec models and speech language models, which can be characterized as follows: 1) The Codec model is inherently designed for information compression; therefore, the training objective of the Codec model aims to preserve as much information as possible within the codebook space to enhance reconstruction. We have discovered that a single channel of codebook is sufficient to reconstruct a significant portion of the audio signal. \textbf{The RVQ structure, in particular, results in the first channel of the codebook containing excessive information.} Consequently, in downstream tasks, whether unconditionally or based on weak conditioning such as \textbf{text}, efficiently generating long and high-quality audio segments remains an unresolved challenge.  2) In order to generate high-quality audio through token modeling with the neural codecs, the rate of discrete representation must be increased, which leads to either exponential growth in codebook size or the generation of long token sequences. Therefore, there is a need for fewer codebooks to achieve this goal.

Based on the findings mentioned above, we attempted to construct a discrete codec model that is more suitable for downstream speech language models. \textbf{Our objective is to include less information in the first channel of the codebook while increasing the missing information on limited channels}. \textbf{We consider that within downstream speech language models, the first-layer quantizer of the Codec model serves as an intermediary module bridging textual input and subsequent quantizers.} By judiciously reducing information within the first-layer quantizer, employing text (which inherently carries less information compared to speech) to generate first Codec(codec in the first quantizer) with lower information content can be more easy. Therefore, we devised the Masked Channel Residual Vector Quantization (MCRVQ) mechanism, which employs the masking mechanism to restrict the quantizers of the first three channels to learn only the compressed audio frame information in the specified space. Simultaneously, We designed a new, more powerful decoder with improved Fourier transform structure \cite{vocos} and attention block. We alos add a complex STFT discriminator \cite{dac} at multiple time-scales \cite{encodec}. Through these modules with enhanced sampling and reconstruction capabilities, Language-Codec achieves excellent reconstruction quality on various test datasets using only four channels, thereby enhancing its compatibility with downstream models. The contributions of Language-Codec are as follows:

\begin{itemize}
    \item \textbf{Language-Codec} is the pioneering discrete \textbf{codec} model formulated from the standpoint of speech \textbf{language} models. Specifically, Language-Codec introduces an innovative MCRVQ structure, which effectively consolidates the information within the codebook.
    \item By utilizing a modern, powerful decoder and multi-scale discriminator. Language-Codec achieves excellent audio reconstruction quality with only \textbf{four} codebook channels.
    \item Language-Codec demonstrates significant outperformance compared to competing audio compression algorithms across various metrics and different test datasets.
    \item The code and pre-trained models of Language-Codec will be open-source. 
\end{itemize}

\section{Related Works}
In recent times, neural acoustic codecs \cite{soundstream,encodec,hificodec,vocos,speechtokenizer,nips2024high,FunCodecAF,ji2024wavtokenizer} have demonstrated remarkable capabilities in reconstructing high-quality audio at extremely low bitrates. Consequently, these codecs have facilitated the application of discrete modeling to a wide range of audio signals, including zero-shot TTS, music generation, and audio generation. Typically, these methods employ an encoder to extract deep features in a latent space, which are subsequently quantized before being fed into the decoder.

To elaborate, Soundstream \cite{soundstream} utilizes a model architecture comprising a fully convolutional encoder-decoder network and a residual vector quantizer (RVQ) to effectively compress speech. Encodec \cite{encodec} employs a streaming encoder-decoder architecture with a quantized latent space, trained in an end-to-end fashion. HiFi-Codec \cite{hificodec} introduces a group-residual vector quantization (GRVQ) technique to reduce the number of quantizers. Vocos \cite{vocos} aims to bridge the gap between time-domain and Fourier-based neural vocoders for high-quality audio synthesis. In order to narrow the gaps between text and acoustic codec tokens, HILCodec~\cite{ahn2024hilcodec}  introduces the
MFBD discriminator to guide codec modeling. APCodec~\cite{ai2024apcodec}  further enhances
reconstruction quality by incorporating ConvNextV2 modules in the encoder and decoder. SpeechTokensizer \cite{speechtokenizer} introduces the concept of using semantic tokens in the first channel of discrete codecs. Recently, both SemantiCodec~\cite{liu2024semanticodec} and DAC~\cite{dac} have demonstrated strong reconstruction performance. SemantiCodec employs a VQ-VAE framework and utilizes a diffusion model to model the VAE's latent space. While it achieves excellent reconstruction at low bitrates, its inference speed is approximately 100 times slower than that of standard codec models. DAC~\cite{dac} incorporates strategies such as factorized codes, L2-normalized codes, and quantizer dropout, making it one of the most advanced codec models.

The primary distinction from the previously mentioned models lies in the fact that, while Language-Codec aims to optimize codec reconstruction performance, it places greater emphasis on the gap between the reconstruction paradigm and downstream generative models. We hypothesize that the first-layer quantizer serves as an intermediary module, connecting the textual input with subsequent quantizers in speech language models. To address this challenge, we propose the MCRVQ mechanism as a solution within the Language-Codec framework.

\begin{figure*}[t]
\centering
\includegraphics[height=6.5cm, width=16cm]{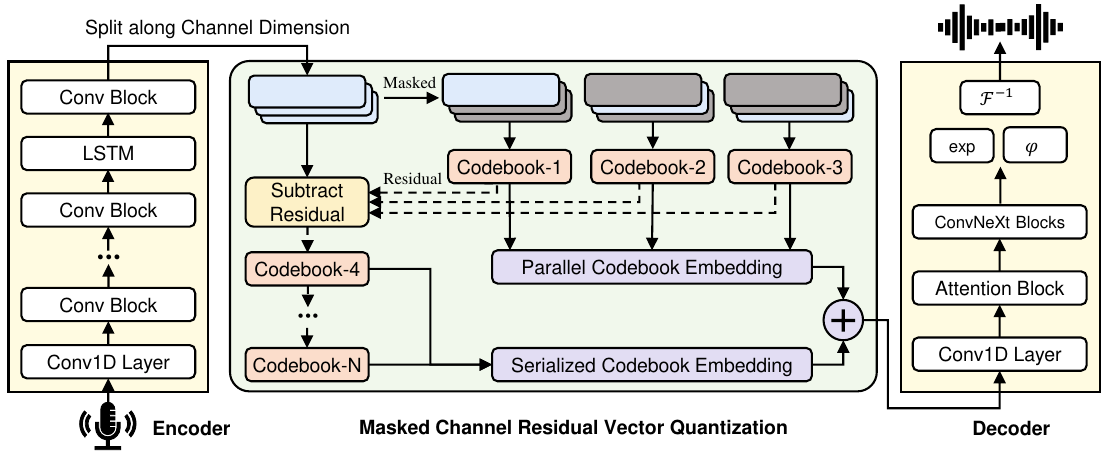}

\caption{The overall architecture for Language-Codec. On the far left is the encoder downsampling module, which still utilizes the model structure of Encodec. On the far right is the decoder upsampling module, where we have replaced it with Vocos' model structure. The middle part is the Masked Channel Residual Vector Quantization module, with the gray blocks indicating the masked portion of temporal information. The dashed lines within the MCRVQ module indicate that the corresponding representations exhibit a decrease in residual values.}
\label{figurejsp1}
\end{figure*}

\section{Language-Codec}
In this section, we will first introduce the overall architecture of Language-Codec, followed by a detailed focus on the encoder module, new decoder module, and Masked Channel Residual Vector Quantization module. Furthermore, we will proceed to elaborate on the specific intricacies of Language-Codec's training processes, with the explicit details of the loss and discriminator.
\subsection{Overall}
As illustrated in Figure \ref{figurejsp1}, the structure of Language-Codec is identical to that of mainstream codec models. It can be divided into three main components: inputting the raw audio signal $X$, and outputting the reconstructed audio signal $\tilde{X}$. It is widely acknowledged that the original single-channel audio signal $X$ is represented as a one-dimensional vector sequence.
\begin{equation}
    X=\left \{ x_{1},x_{2},\cdots ,x_{T}  \right \},T=d*sr 
\end{equation}
Where $sr$ is the audio sample rate and $d$ is the audio duration. Language-Codec passes the raw audio $X$ through three modules. 1) an encoder network that takes the input audio and generates a latent feature representation $Z$; 2) a combination of parallel and serialized quantization layer $q$ that produces a compressed representation $Z_{q}$; and 3) a decoder that reconstructs the audio signal $\tilde{X}$ from the compressed latent representation $Z_{q}$. The model is trained end-to-end, optimizing a reconstruction loss applied over both time and frequency domains, along with a perceptual loss in the form of discriminators operating at different resolutions. 
\subsection{Encoder and Decoder}
Follow Encodec \cite{encodec}, the encoder model consists of a 1D convolution with $C$ channels and a kernel size of 7 followed by $B$ convolution blocks. Each convolution block is composed of a single residual unit followed by a down-sampling layer consisting of a stridden convolution, with a kernel size of twice the stride $S$. The residual unit contains two convolutions with kernel size 3 and a skip-connection. The number of channels is doubled whenever down-sampling occurs. The convolution blocks are followed by a two-layer LSTM for sequence modeling and a final 1D convolution layer with a kernel size of 7 and $D$ output channels. Following Encodec \cite{encodec}, we use $C$ = 32, $B$ = 4, and (2, 4, 5, 8) as $S$. We use ELU as a non-linear activation function. With this setup, Language-Codec outputs 75 latent steps per second of audio at 24 kHz.

Language-Codec does not employ a mirrored decoder upsampling structure. The standard practice involves using a stack of dilated convolutions to increase the receptive field, and transposed convolutions to sequentially upsample the feature sequence to the
waveform.  However, this design is known to be susceptible to aliasing artifactsInstead, following Vocos \cite{vocos}, we maintain consistent feature resolution at all depths, achieving waveform upsampling through inverse Fourier transform. In the decoder section, the target audio signal $\tilde{X}$ is represented using Short-Time Fourier Transform (STFT):
\begin{equation}
    STFT(\tilde{X}_{\left [ m,k \right ] } )=\sum_{n=0}^{N} \tilde {X}\left [ n \right ] w\left [ n-m \right]e^{-j2\pi kn/K}
\end{equation}
Here, $K$ represents the number of frequency points after performing the Discrete Fourier Transform (DFT), while $k$ denotes the frequency index. $N$ corresponds to the number of points in the sampled sequence, with $n$ representing a particular sample point, and $m$ indicating the index length. In the practical implementation, the Short-Time Fourier Transform (STFT) is performed by applying a series of Fast Fourier Transforms (FFTs) to overlapping and windowed frames of data. The window function advances or hops through time to create these frames. 

Therefore, for the representation of the intermediate signals $Z_{q}$ after quantization, the Language-Codec only needs to input $Z_{q}$ into the conv1D layer, attention block, ConvNeXt \cite{convnext} blocks, which serves as the fundamental backbone. Subsequently, a Fourier transform is performed on the real-valued signals. Notably, we introduced an attention module in the decoder to enhance the sequence modeling capability of the upsampling module. During experiments, we observed that although 3-second audio segments were randomly selected during training, there were no issues with length extrapolation when reconstructing longer audio segments during inference. In ConvNeXt Block, it first embeds the input features into a hidden dimensionality and then applies a sequence of convolutional blocks. Each block is composed of a large-kernel-sized depthwise convolution, followed by an inverted bottleneck that projects features into a higher dimensionality using pointwise convolution. GELU (Gaussian Error Linear Unit) activations are used within the bottleneck, and Layer Normalization is employed between the blocks. Regarding the transformation of real-valued signals, we utilize a single side band spectrum, resulting in $n_{fft}/2 + 1$ coefficients per frame. Since we parameterize the model to output both phase and magnitude values, the activations of the hidden dimensions are projected into a tensor $h$ with $n_{fft} + 2$ channels and subsequently split into:
\begin{equation}
    q=h\left [1 : n_{fft}/2 + 1 \right ] ;p= h\left [n_{fft}/2 + 2 : n  \right ] 
\end{equation}
where $q$ stands for magnitude, $p$ stands for argument, Finally, we represent complex-valued coefficients as:
\begin{equation}
    STFT =exp(q) \cdot (\cos p + j\sin p)
\end{equation}
Finally, the inverse Fourier transform $\mathcal{F} ^{-1} $ can be used to reconstruct the final audio.
\subsection{Masked Channel Residual Vector Quantization}
Within the Masked Channel Residual Vector Quantization module, our aim is to minimize the informational content in the initial channel of the codebook, while augmenting the amount of information compensated on \textbf{constrained} channels. To achieve this objective, a hybrid structure combining parallel and serial quantization is employed within the Language-codec framework. In the initial $N_{q}$ layers, each quantizer independently processes a segment of the original information compressed at the base layer, concurrently producing corresponding codebooks and embedding vectors. For the subsequent layers, extending from $N_{q}$ to $N$, each quantizer sequentially subtracts the embedding vectors generated by all preceding quantizers, utilizing this resultant as the input for the current quantization process.

More specifically, concerning the quantizers operating in parallel within the initial $N_{q}$ layers, the Language-Codec introduces the Masked Channel mechanism to achieve mean quantization of the latent space information $Z$ on the first $N_{q}$ channels of the quantizer. In the actual training process, we simply set $N_{q}$ to 3. We divide the compressed audio frame into $N_{q}$ equal parts and use $M$ to represent the portion to be masked and $\bar{M} $ to represent the remaining portion. Following the order of the quantizers, we mask the specified portion of the quantizer and retain $\frac{1}{N_{q}} $ of the latent space information $Z$, which is then directly fed into the quantizer. Therefore, when the quantizer generates $i$ the intermediate result $\hat{Z_{i}}$ for the layer $i$ $(1 \le i \le N_{q} )$, this process can be represented by the following equation:
\begin{equation}
    P(\hat{Z_{i}} |\bar{M}Z )=P(\hat{Z_{i}} |(1-M)Z )=P(\hat{Z_{i}}|\frac{Z}{N_{q}})  
\end{equation}

For the quantizers after the $N_{q}$ channels, we still retain the information $\hat{Z}_{j}$ obtained by subtracting the residual of $Z$ from the previous $N_{j}$ channels, and then feed it into the quantizer $j$ $(N_{q}+1 \le j\le N )$. It is noteworthy that, given the parallel architecture of the quantizers in the first $N_{q}$ layers, the input to the quantizer at the $N_{q}+1$ layer must sequentially subtract the representations of each preceding layer, rather than being simply denoted as $Z-\hat{Z}_{j}$. The generation process for the $N_{q}+1$ layer can be distinctly represented as follows: 
\begin{equation}
    P(\hat{Z}_{N_{q}+1} |Z-\sum_{i=1}^{N_{q}}\hat{Z}_{i})
\end{equation}

After passing through $N$ quantizers, the information on each channel is fused to obtain the final result $Z_{q}$. The function of the fusion layer is to concatenate the output embedding matrices from the parallel quantizers and the serial quantizers along the channel dimension. A similar fusion operation is applied to the codebook vectors as well. In summary, The Masked Channel Residual Vector Quantization mechanism can be represented as follows:
\begin{equation}
    \begin{aligned}
    P(Z_{q}|Z)&=\prod_{i=1}^{N_{q}} P(\hat{Z_{i}} |\bar{M}Z )P(\hat{Z}_{N_{q}+1} |Z-\sum_{i=1}^{N_{q}}\hat{Z}_{i})\\
    &\times\prod_{j=N_{q}+1}^{N-1} P(\hat{Z}_{j+1}|Z-\hat{Z}_{j})
    \end{aligned}
\end{equation}
\subsection{Discriminator and Loss}
The adversarial loss is used to promote perceptual quality. We employ the multi-period discriminator (MPD) as defined by \cite{mpd}
and multi-resolution discriminator (MRD) \cite{mrd}. Furthermore, to learn discriminative features about a specific sub-band and provide a stronger gradient signal to the generator, following \cite{nips2024high}, we use a multi-scale discriminator (MSD) and a complex STFT discriminator \cite{soundstream} at multiple time-scales \cite{encodec}. We adopt a hinge loss formulation instead of the least squares GAN objective, as suggested by \cite{soundstream}. To train the discriminator, we can optimize the following objective function $\mathcal{L}_{dis}(X,\tilde{X} )$:
\begin{equation}
    \frac{1}{K} \sum_{k=1}^{K} max (0, 1-D_{k}(X)) + max (0, 1+D_{k}(\tilde{X}))
\end{equation}
The variable $K$ represents the number of discriminators. $D_{k}$ represents the $k$-th discriminator. Regarding the loss for the generator, the Language-Codec model consists of four components: quantizer loss, mel-spectrum reconstruction loss, adversarial loss, and feature matching loss. The quantizer loss can be defined as follows:
\begin{equation}
    \mathcal{L}_{q}(Z,Z_{q} ) =\sum_{i=1}^{N} \left \| Z_{i}-\hat{Z}_{i}  \right \| _{2}^{2} 
\end{equation}
The mel-spectrum reconstruction loss can be defined as follows:
\begin{equation}
    \mathcal{L}_{mel}(X,\tilde{X}  ) = \left \| Mel(X)-Mel(\tilde{X})  \right \| _{1}  
\end{equation}
Furthermore, we can define the adversarial loss as a hinge loss over the logits of these discriminators:
\begin{equation}
    \mathcal{L}_{adv}  = \frac{1}{K} \sum_{k=1}^{K} max (0, 1-D_{k}(\tilde{X} ))
\end{equation}
The feature matching loss, denoted as $\mathcal{L}_{feat}$ , is calculated as the mean of the distances between the $l$th feature maps of the $k$th subdistriminator:
\begin{equation}
    \mathcal{L}_{feat} =\frac{1}{K*L}\sum_{k}\sum_{l}\left \| D_{k}^{l}(X)-D_{k}^{l}(\tilde{X})  \right \| _{1}
\end{equation}
In the end, the total loss of the generator $\mathcal{L}_{gen}$ is:
\begin{equation}
\mathcal{L}_{gen}=\lambda_{q}\mathcal{L}_{q}+\lambda _{mel}\mathcal{L}_{mel} +\lambda_{adv}\mathcal{L}_{adv}+\lambda _{feat}\mathcal{L}_{feat}   
\end{equation}
where $\lambda_{q}$, $\lambda_{mel}$, $\lambda_{adv}$, $\lambda_{feat}$ are the hyper-parameters to control the training
objective function.

\section{Experiments}
\subsection{Experiment Setup}
\textbf{Datasets.}  
Language-Codec was trained on a comprehensive 50,000-hour speech dataset (We verify the effect of training dataset of different sizes on the model in Appendix \ref{appendix training data}). We employed a combination of Librilight's small and medium collections \cite{librilight}, speech segments from DNS Challenge 4 \cite{dns4}, the Common Voice dataset (version 16.0) \cite{commonvoice}, LibriTTS \cite{librispeech} training set, and 20,000 hours of internal Chinese data as the integrated training dataset. To ensure a fair comparison of codec models' performance, we conducted inference testing on the LibriTTS \cite{LibriTTS} Test-Clean and Test-Other sets to evaluate codecs' restoration effectiveness in common and noisy environments respectively. Additionally, we performed tests on the LJSpeech dataset to simulate out-of-domain scenarios. For downstream speech language models, we utilized the LibriTTS training set to train zero-shot text-to-speech models. Inference testing was carried out on the LibriSpeech \cite{librispeech} Test-Clean sets, following VALL-E \cite{valle} and MobileSpeech \cite{ji2024mobilespeech}, we filtered audio samples of 4-10 seconds from the LibriSpeech Test-Clean sets.\\
\textbf{Automatic metrics.}
For objective evaluation of our codec models, we employ the UTMOS \cite{utmos} automatic Mean Opinion Score (MOS) prediction system. UTMOS can yield scores highly correlated with human evaluations and is restricted to 16 kHz sample rate. we also adopt the metrics from speech enhancement fields, such as the PESQ \cite{pesq}, STOI, and the F1 score for voiced/unvoiced classification (V/UV F1), following the methodology proposed by Vocos\cite{vocos} to evaluate the performance of discrete codecs. Moreover, we have aligned all our zero-shot TTS experiments metrics with VALL-E \cite{valle}. To evaluate speaker similarity (SPK) between the original prompt and synthesized speech, we employ WavLM-TDNN \cite{wavlm}. However, due to updates in the repository, we have updated the feature extractor in WavLM, but all our models have been tested by using the same metrics. For assessing automatic speech recognition (ASR) performance, we conduct ASR on the generated audio and calculate the word error rate (WER) compared to the original transcriptions. In this experiment, we utilize the HuBERT-Large \cite{hubert} model fine-tuned on LibriSpeech 960h as the ASR model. This model is a CTC-based model without language model fusion.\\
Due to space constraints, details of the Baselines and Training and Inference Settings are in Appendix \ref{appendix baselines} and Appendix \ref{appendix training}. Subjective evaluations are detailed in Appendix \ref{appendix human}.
\begin{table*}[t]
\centering
\caption{The results of different codec models on the LibriTTS Test-Clean and Test-Other dataset. \textbf{Since downstream generative models use at most 8 layers of quantizers, we limit our comparisons to the 6 kbps and 3 kbps dimensions, even though 12 kbps can achieve better reconstruction performance.}}
\begin{adjustbox}{width=\textwidth}
\begin{tabular}{cccccccc}
\hline
\textbf{Model} & \textbf{Bandwidth} & \textbf{Nq}  & \textbf{UTMOS $\uparrow$}   & \textbf{PESQ $\uparrow$} & \textbf{STOI $\uparrow$} &\textbf{ V/UV F1 $\uparrow$} & \textbf{SPK $\uparrow$} \\
\hline
LibriTTS Test-Clean sets\\
\hline
GT & - & - & 4.0562 &- &- & - & -\\ 
Opus &6.0kbps	&-	&2.7961&	2.5860	&0.9367&	0.9408	&0.7701 \\
EVS &7.2kbps&	-	&3.4539 &	3.0988&	0.9317&	0.9453	&0.8524\\
Encodec	&3.0kbps	&4	&2.3070 &	2.0517&	0.9007&	0.9198	&0.7860\\
Encodec	&6.0kbps&	8	&3.0399&	2.7202&	0.9391	&0.9527	&0.8822\\
Vocos &3.0kbps	&4	 &3.5390 &2.4026 & 0.9231 & 0.9358 &0.7892 \\			
Vocos &	6.0kbps	& 8	& 3.6954 & 2.8069 & 0.9426 & 0.9437 &0.8608\\
SpeechTokenizer &	3.0kbps	&4	& 3.5632	&1.9311&	0.8778&	0.9273	&0.6587\\
SpeechTokenizer &6.0kpbs	&8	&3.8794&	2.6121&	0.9165	&0.9495	&0.8311\\
DAC &	3.0kbps	& 4	&2.9902	&2.4091	&0.9118&	0.9531	&0.8129\\
DAC	&6.0kbps & 8	&3.6804&3.5558&	0.9257	&0.9711	&0.8715\\
Language-Codec &	3.0kbps&	4&	\textbf{3.7875}	& \textbf{3.2675}	& \textbf{0.9493}	& \textbf{0.9657} &	\textbf{0.8698}\\
Language-Codec &	6.0kbps	&8	& \textbf{4.0372}	&\textbf{3.8813}&	\textbf{0.9715}	& \textbf{0.9754}	& \textbf{0.9318}\\
\hline
\end{tabular}
\end{adjustbox}
\label{table1}
\end{table*}

\subsection{Reconstruction Evaluation}
We evaluated the reconstruction performance of the codec model on the LibriTTS Test-Clean dataset \cite{LibriTTS}.  Considering that the primary purpose of the discrete codecs is to serve as an audio representation for downstream tasks, excessive channel numbers would significantly burden downstream speech language models. Therefore, we conducted a comparison between four-channel and eight-channel dimensions. \textbf{We believe that the results on the LibriTTS Test-Clean subset adequately reflect the reconstruction performance of the codec model.} To further evaluate the reconstruction performance of the codec model in noisy environments and out-of-domain scenarios, we additionally compare the performance of different codec models on the LibriTTS Test-Other~\cite{LibriTTS} and LJSpeech datasets in Appendix \ref{appendix generalization}.



Based on the observations from Table \ref{table1}, the following conclusions can be drawn: 1) Regarding the audio reconstruction of the four-channel codecs, the Language-Codec model significantly outperforms all baseline models in terms of objective metrics. While there is a slight decrease in audio reconstruction quality when the number of channels is reduced from eight to four in the baseline models, the Language-Codec model maintains a consistently good reconstruction performance. Additionally, it is noteworthy that the four-channel reconstruction of Language-Codec even surpasses the eight-channel performance of several baseline models. For instance, in terms of the PESQ and STOI metrics, the four-channel Language-Codec model outperforms the eight-channel SpeechTokenizer model by 0.6 and 0.03 in LibriTTS Test-Clean sets. Furthermore, in the UTMOS metric, the four-channel Language-Codec model significantly outperforms the eight-channel Encodec model. 2) In the eight-channel codecs reconstruction, the Language-Codec model also maintains SOTA reconstruction quality. Although the eight-channel SpeechTokenizer model achieved similar scores to the Language-Codec model in terms of the UTMOS metric, it significantly underperformed in other metrics such as SOTI, SPK, and PESQ compared to the Language-Codec model, and even performed noticeably worse than the Encodec model. Considering the overall auditory perception and average audio quality, the Language-Codec model achieves the best performance.

\subsection{Zero-shot TTS Evaluation}
Regarding the inference phase, to ensure fair comparisons, we followed the experimental protocols outlined in VALL-E \cite{valle} and employed the LibriSpeech Test-Clean dataset \cite{librispeech}, ensuring no overlap with our training data. We specifically utilized samples from the LibriSpeech Test-Clean dataset with durations ranging from 4 to 10 seconds, resulting in a subset of 2.2 hours. Following VALL-E, we use the whole transcription and the first 3 seconds of the utterance as the phoneme and acoustic prompts respectively, and ask the model to generate the continuations.
\begin{table*}[t]
\centering
\caption{Evaluation of zero-shot TTS models with different codecs on the LibriSpeech Test-Clean corpus.}
\begin{adjustbox}{width=\textwidth}
\begin{tabular}{cccccc}
\hline
\textbf{Zero-shot TTS Model} & \textbf{WER $\downarrow$} & \textbf{SPK $\uparrow$} & \textbf{MOS-Q $\uparrow$} & \textbf{MOS-P $\uparrow$ } & \textbf{MOS-S $\uparrow$ } \\
\hline
VALL-E w/ Encodec & 4.3 & 0.6115 &3.73$\pm$0.09 & 3.76$\pm$0.12 & 3.74$\pm$0.11\\ 
VALL-E w/ Encodec+Vocos & 4.3 & 0.6198 & 3.83$\pm$0.07 & 3.81$\pm$0.13 & 3.82$\pm$0.06\\ 
VALL-E w/ Language-Codec & \textbf{3.8} & \textbf{0.6995} & \textbf{4.01$\pm$0.09} & \textbf{3.89$\pm$0.12} & \textbf{3.99$\pm$0.08}\\ 
\hline
MobileSpeech w/ Encodec &3.2&0.6776 &3.91$\pm$0.10 & 3.99$\pm$0.12 & 3.98$\pm$0.11\\ 
Mobilespeech w/ Encodec+Vocos & 3.1 &0.6883 &4.05$\pm$0.08 & 4.03$\pm$0.12 & 4.04$\pm$0.09\\ 
MobileSpeech w/ Language-Codec & \textbf{2.9} & \textbf{0.7712} &\textbf{4.20$\pm$0.11} & \textbf{4.09$\pm$0.07} & \textbf{4.18$\pm$0.10}\\ 
\hline
\end{tabular}
\end{adjustbox}
\label{table2}
\end{table*}
Given the widespread adoption of Encodec models in downstream speech language models for extracting features as intermediate acoustic tokens, we have selected it as the primary baseline model in our study. For the Vocos version, we maintain the practice of training downstream models using feature representations extracted by the Encodec model. However, during the inference and decoding phase, we replace the Encodec decoder with the Vocos decoder. A more comprehensive explanation of this approach, including the rationale and methodology behind it, will be provided in the subsequent section dedicated to ablation experiments. Additionally, in the case of the Language-Codec version, we focus exclusively on training models with codec representations derived from the Language-Codec model. 


As shown in Table \ref{table2}, the experimental results indicate that different discrete codec representations do not exhibit significant differences in terms of the Word Error Rate (WER) metric. However, for the Speaker Similarity (SPK) metric, we observe that the codecs extracted by the Language-Codec model perform better on downstream models. By merely replacing the codec representation, the average speaker similarity increases by 10$\%$-15$\%$. Additionally, in subjective Mean Opinion Score (MOS) evaluations, we discover that the codec representations extracted by the Language-Codec model exhibit certain improvements in terms of audio quality and audio similarity compared to those extracted by the encoder model. However, no significant differences are observed in terms of prosodic representations. Moreover, during the training process of our downstream zero-shot Text-to-Speech model, we find that when the downstream model predicts codecs generated by the Language-Codec model, the accuracy of codec prediction decreases when the number of channels exceeds four. Although this does not have a significant impact on the performance of the downstream model, future endeavors could explore the use of smaller or variable codebooks to further enhance the results.

\subsection{Ablation experiment}
In this section, we conducted a detailed analysis of MCRVQ module. The ablation experiments regarding the new decoder and the multi-scale discriminator are presented in Appendix \ref{appendix decoder} and \ref{appendix discriminator}.

We validated the role of the Masked Channel Residual Vector Quantization (MCRVQ) module in the language-codec model. Considering that the design purpose of the MCRVQ mechanism is to reduce the difficulty of text generation in downstream tasks, we conducted ablation experiments on the zero-shot TTS model downstream. Specifically, we first replaced the MCRVQ module with the RVQ module (eight quantizers) while keeping the same training steps and other configurations. We refer to this experiment setup as Language-Codec w/o MCRVQ. We used Language-Codec w/o MCRVQ to extract the corresponding discrete codec features and retrained the downstream VALL-E and MobileSpeech models. The experimental results, as shown in Table \ref{table4}, revealed that there was no significant difference (0.2) between Language-Codec w/o MCRVQ and Language-Codec in terms of the robustness metric WER. However, in terms of the objective metric of speaker similarity, omitting the MCRVQ module resulted in a decrease of 0.06 similarity in VALL-E and MobileSpeech, respectively, indicating that the MCRVQ module indeed enhances the codec generation capability of the downstream speech synthesis model by weakening the difficulty of text generation for codec.

In addition, we also conducted corresponding subjective CMOS tests. From Table \ref{table5}, it can be observed that in the autoregressive discrete codec modeling experiments of the VALL-E model, the CMOS values of the synthesized audio decreased by 0.19 when the MCRVQ module was omitted compared to the original Language-Codec model. Similarly, in the parallel discrete codec modeling experiments of the MobileSpeech model, the CMOS values of the synthesized audio decreased by 0.25 when the MCRVQ module was omitted compared to the original Language-Codec model, which further indicated that the codec generated by the Language-Codec w/o MCRVQ model had lower subjective audio quality and audio similarity than the codec generated by the Language-Codec.

\begin{table}
\centering
\caption{The ablation experiments of the MCRVQ module, we assessed the performance of WER and SPK metrics.}
\begin{adjustbox}{width=0.48\textwidth}
\begin{tabular}{cccc}
\hline
\textbf{Model}  & \textbf{Codec Model}   & \textbf{WER $\downarrow$} & \textbf{SPK $\uparrow$} \\
\hline
VALL-E & Language-Codec w/o MCRVQ & 4.1 & 0.6383 \\ 
VALL-E & Language-Codec & \textbf{3.8} & \textbf{0.6995}\\ 
MobileSpeech& Language-Codec w/o MCRVQ& 3.1 & 0.7103 \\ 
MobileSpeech& Language-Codec & \textbf{2.9} & \textbf{0.7712} \\ 
\hline
\end{tabular}
\end{adjustbox}
\label{table4}
\end{table}

\begin{table}
\centering
\caption{The ablation experiments of the MCRVQ module, we assessed the performance of subjective metric CMOS.}
\begin{adjustbox}{width=0.48\textwidth}
\begin{tabular}{ccc}
\hline
\textbf{Model}  & \textbf{Codec Model}   & \textbf{CMOS $\uparrow$} \\
\hline
VALL-E & Language-Codec w/o MCRVQ & -0.19 \\ 
VALL-E & Language-Codec & 0.00 \\ 
MobileSpeech& Language-Codec w/o MCRVQ& -0.25 \\ 
MobileSpeech& Language-Codec & 0.00 \\ 
\hline
\end{tabular}
\end{adjustbox}
\label{table5}
\end{table}

\section{Conclusions}
In this article, we propose Language-Codec, a discrete acoustic codec model that enhances adaptation to downstream speech language models. Through improved model architecture and a unique Masked Channel residual vector quantization mechanism, we achieve excellent audio reconstruction quality with just four layers of codecs. The Language-Codec model demonstrates effective audio restoration performance in both clean audio and noisy environments. Furthermore, we validate the generalization capability of Language-Codec in unseen domains and its generation ability in downstream zero-shot TTS models, yielding satisfactory results. We envision Language-Codec as a state-of-the-art foundational codec model for future research in the field of speech generation.

\section{Acknowledgments}
 This work was supported in part by the National Natural Science Foundation of China under Grant No.62222211 and No.U24A20326

\section{Limitations}
While Language-Codec models demonstrate superior reconstruction quality compared to some notable comparative models, it is important to note that due to time constraints, Language-Codec models have been trained solely on speech corpora and validated solely on downstream speech language models. Although the majority of codec models also exclusively support speech data, we aspire to develop a more comprehensive and versatile codec model. In the future, we plan to incorporate a larger training dataset (consisting of several hundred thousand hours) to encompass a wider range of signal types, including audio, music, and more.

\bibliography{acl_latex}

\appendix


\begin{table*}
\centering
\caption{The reconstruction results of different training datasets on LibriTTS Test-Clean corpus.}
\begin{adjustbox}{width=\textwidth}
\begin{tabular}{cccccccc}
\hline
\textbf{Model} & \textbf{Dataset} & \textbf{Nq}  & \textbf{UTMOS $\uparrow$}   & \textbf{PESQ $\uparrow$} & \textbf{STOI $\uparrow$} &\textbf{ VUV F1 $\uparrow$} & \textbf{SPK $\uparrow$} \\
\hline
Language-Codec	& LibriTTS (585 hours) &4	&\textbf{3.7901} & 3.2652&0.9495& 0.9652 & 0.8688\\
Language-Codec	& LibriTTS (585 hours) &	8	& \textbf{4.0402} &3.8581 &	0.9701	& 0.9711	& 0.9165\\
Language-Codec & Paper w/o 20k internal data	&4	 & 3.7867 & 3.2598 & \textbf{0.9499}  & 0.9654  & 0.8688 \\			
Language-Codec & Paper w/o 20k internal data	& 8	& 4.0279 & 3.8645 & \textbf{0.9716} & 0.9735 & 0.9268 \\
Language-Codec & 50k hours&	4&	3.7875	& \textbf{3.2675}	& 0.9493	& \textbf{0.9657} &	\textbf{0.8698}\\
Language-Codec & 50k hours&8	& 4.0372	&\textbf{3.8813}&	0.9715	& \textbf{0.9754}	& \textbf{0.9318}\\
\hline
\end{tabular}
\end{adjustbox}
\label{table10}
\end{table*}

\begin{table*}[t]
\centering
\caption{The reconstruction results of different codec models on the LibriTTS Test-Other dataset (noisy environment).}
\begin{adjustbox}{width=\textwidth}
\begin{tabular}{cccccccc}
\hline
\textbf{Model} & \textbf{Bandwidth} & \textbf{Nq}  & \textbf{UTMOS $\uparrow$}   & \textbf{PESQ $\uparrow$} & \textbf{STOI $\uparrow$} &\textbf{ V/UV F1 $\uparrow$} & \textbf{SPK $\uparrow$} \\
\hline
LibriTTS Test-Other sets\\
\hline
GT & - & - & 3.4831 &- &- & - & -\\ 
Opus &6.0kbps	&-	&2.2628&	2.5701	&0.9233&	0.9265	&0.7563 \\
EVS &7.2kbps&	-	&2.8845	& 2.8456 &	0.9102&	0.9256	&0.8407\\
Encodec	&3.0kbps	&4	&2.0883 &	2.0529&	0.8835&	0.8926	&0.7724\\
Encodec	&6.0kbps&	8	&2.6568&	2.6818&	0.9241	&0.9338	&0.8763\\
Vocos &3.0kbps	&4	&3.0558 &2.1933 &0.8967 & 0.9051 &0.7592\\			
Vocos &	6.0kbps	& 8	 &3.1956 & 2.5590 & 0.9209 & 0.9202 & 0.8363\\
SpeechTokenizer &	3.0kbps	&4	&3.0183	&1.7373&	0.8371&	0.8907	&0.6071\\
SpeechTokenizer &6.0kpbs	&8	&3.2851&	2.3269&	0.8811	&0.9205	&0.7925\\
DAC &	3.0kbps	&4	& 2.5981	&2.2380	&0.8869&	0.9361	&0.7947\\
DAC	&6.0kbps	&8	&3.1701& 3.2874&	0.9341	&0.9578	&0.8563\\
Language-Codec &	3.0kbps&	4&\textbf{3.2218}	&\textbf{2.9844}	&\textbf{0.9207}	&\textbf{0.9483}& \textbf{0.8466}	\\
Language-Codec &	6.0kbps	&8	& \textbf{3.4801}	&\textbf{3.7218}& \textbf{0.9498}	&\textbf{0.9654}	&\textbf{0.9153}\\
\hline
\end{tabular}
\end{adjustbox}
\label{table_testother}
\end{table*}

\begin{table*}[t]
\centering
\caption{The reconstruction results of different codec models on the LJSpeech dataset (out domain scenarios).}
\begin{adjustbox}{width=\textwidth}
\begin{tabular}{cccccccc}
\hline
\textbf{Model} & \textbf{Bandwidth} & \textbf{Nq}  & \textbf{UTMOS $\uparrow$}   & \textbf{PESQ $\uparrow$} & \textbf{STOI $\uparrow$} &\textbf{ V/UV F1 $\uparrow$} & \textbf{SPK $\uparrow$} \\
\hline
LJSpeech\\
\hline
GT & - & - & 4.3794 &- &- & - & -\\ 
Opus &6.0kbps	&-	&2.7640	&2.1433	&0.1245& 0.9489& 0.7098\\
EVS &7.2kbps&	-	&	3.8991&3.0560&	0.9507&		0.9521&0.8551\\
Lyra-v2 &	3.2kbps	&-	&	3.3773&	2.4182&0.9161&	0.9421	&0.7041\\
Lyra-v2	&6.0kbps	&-	&3.9591&2.8853&0.9418	&0.9551&0.8007\\
Encodec	&3.0kbps	&4	& 2.3905&2.0194	&0.9058	&0.9326		&0.8177\\
Encodec	&6.0kbps&	8	&3.2286&2.6633	&0.9441	& 0.9555	& 0.8952\\
Vocos &3.0kbps	&4	 & 3.7880&2.5006&0.9310&0.9388 &0.7801 \\			
Vocos &	6.0kbps	& 8	&  4.0332&2.9258 &0.9497  &  0.9459&0.8339\\
SpeechTokenizer &	3.0kbps	&4	& 3.9908	&2.0458&0.9021&	0.9299&0.6793\\
SpeechTokenizer &6.0kpbs &8& 4.2373 &2.6413&0.9316		&0.9452&0.8332\\
Language-Codec &	3.0kbps&	4&	\textbf{4.1416}	& \textbf{3.2488} & \textbf{0.9493}	& \textbf{0.9612} & \textbf{0.8675}\\
Language-Codec &	6.0kbps	&8	& \textbf{4.3561}	& \textbf{3.7456}&	\textbf{0.9704}	& \textbf{0.9732} & \textbf{0.9389} \\
\hline
\end{tabular}
\end{adjustbox}
\label{table3}
\end{table*}

\begin{table}[htbp]
\centering
\caption{The ablation experiment of multi-scale discrimination (msdm) on the LibriTTS Test-Clean corpus.}
\begin{adjustbox}{width=0.48\textwidth}
\begin{tabular}{cccccc}
\hline
\textbf{Model}  & \textbf{UTMOS $\uparrow$}   & \textbf{PESQ $\uparrow$} & \textbf{STOI $\uparrow$} &\textbf{ V/UV F1 $\uparrow$} \\
\hline
Language-Codec w/o msdm & 3.8682	&3.5117 & 0.9636 & 0.9682 	\\
Language-Codec & \textbf{4.0372}	&\textbf{3.8813}&	\textbf{0.9715}	& \textbf{0.9754}	\\
\hline
\end{tabular}
\end{adjustbox}
\label{table8}
\end{table}

\begin{table*}[htbp]
\centering
\caption{The ablation experiment of decoder on the LibriTTS Test-Clean corpus.}
\begin{adjustbox}{width=\textwidth}
\begin{tabular}{cccccc}
\hline
\textbf{Model}  & \textbf{UTMOS $\uparrow$}   & \textbf{PESQ $\uparrow$} & \textbf{STOI $\uparrow$} &\textbf{ V/UV F1 $\uparrow$} & \textbf{SPK $\uparrow$} \\
\hline
Language-Codec w/o decoder & 3.2557	&3.0654 & 0.9456 & 0.9611 	& 0.9096  \\
Language-Codec w/o attention block & 3.8719	&3.6225 & 0.9642 & 0.9688 	& 0.9208\\
Language-Codec & \textbf{4.0372}	&\textbf{3.8813}&	\textbf{0.9715}	& \textbf{0.9754}	& \textbf{0.9318}\\
\hline
\end{tabular}
\end{adjustbox}
\label{table9}
\end{table*}

\begin{table*}[t]
\centering
\caption{Evaluation of mobilespeech with different quantizers on the LibriSpeech Test-Clean corpus.}
\begin{adjustbox}{width=\textwidth}
\begin{tabular}{ccccccc}
\hline
\textbf{Zero-shot TTS Model}& \textbf{Nq $\downarrow$} & \textbf{WER $\downarrow$} & \textbf{SPK $\uparrow$} & \textbf{MOS-Q $\uparrow$} & \textbf{MOS-P $\uparrow$ } & \textbf{MOS-S $\uparrow$ } \\
\hline
MobileSpeech w/ Language-Codec &4& 3.2 &0.7254 &4.04$\pm$0.09 & 4.06$\pm$0.12 & 4.11$\pm$0.11\\  
MobileSpeech w/ Language-Codec & 8& \textbf{2.9} & \textbf{0.7712} &\textbf{4.20$\pm$0.11} & \textbf{4.09$\pm$0.07} & \textbf{4.18$\pm$0.10}\\ 
\hline
\end{tabular}
\end{adjustbox}
\label{table11}
\end{table*}


\begin{figure*}[htbp]
\centering
\includegraphics[height=5.5cm, width=16cm]{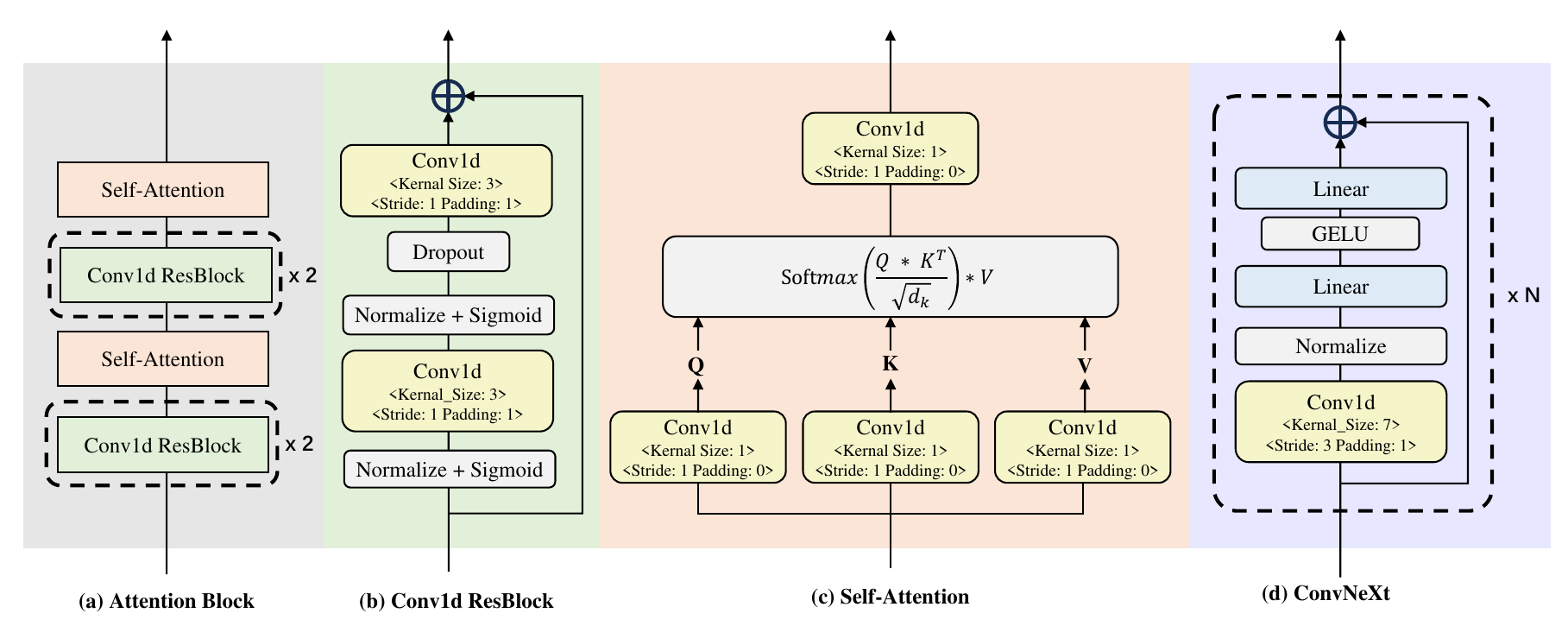}

\caption{The overall architecture of Attention Block and ConvNeXt Blocks inside Decoder. Subfigures (b) and (c) show the more fundamental structure in the Attention Block. The text surrounded by “$<>$” indicates the parameter settings of Conv1d.}
\label{figurejsp_appendix}
\end{figure*}

\section{Ablation experiments about varied training data volumes }

\label{appendix training data}

As shown in Table \ref{table10}. We found that the size
of the dataset does not significantly impact speech reconstruction. Particularly noteworthy is the internal dataset comprising 20k
hours, which, given its Chinese language corpus, exerts minimal
influence on the results presented in the paper. Dataset size may
more influence generalization (more languages) instead of reconstruction. 

\section{Baselines}
\label{appendix baselines}
Following Encodec \cite{encodec}, we considered several traditional speech compression models as baselines. Opus \cite{opus} is a versatile speech and audio codec model that was standardized by the IETF in 2012. EVS \cite{evs} is a codec standardized by 3GPP in 2014 and specifically developed for Voice over LTE (VoLTE). We also utilized the official implementation available in Lyra2 \footnote{\url{https://github.com/google/lyra}} at bit rates of 3.2 kbps and 6 kbps. Additionally, we selected three state-of-the-art codec models based on RVQ as baselines. To ensure a fair comparison, we employed the official weight files provided by the Encodec \footnote{\url{https://github.com/facebookresearch/encodec}} \cite{encodec} , Vocos \footnote{\url{https://github.com/gemelo-ai/vocos}} \cite{vocos} , SpeechTokenizer \footnote{\url{https://github.com/ZhangXInFD/SpeechTokenizer}} \cite{speechtokenizer} and DAC \footnote{\url{https://github.com/descriptinc/descript-audio-codec}} \cite{dac} frameworks. For downstream speech language models, we replicated two zero-shot TTS models based on discrete codecs modeling: VALL-E \cite{valle} , representing autoregressive modeling, and MobileSpeech \cite{ji2024mobilespeech}, representing fully parallel modeling.
\section{Training and Inference Settings}
\label{appendix training}
We train Language-Codec up to 2 million iterations, with 1 million iterations allocated to both the generator and discriminator on 8 NVIDIA A100 40G GPUs. Throughout the entire training process, all input speech samples were resampled to 24 kHz, and the batch size was 100. During the training phase, we uniformly truncated excessively long segments in the training data to a fixed length of 10 seconds and subsequently performed a random crop of the waveform to obtain audio snippets of 1-second duration for feeding Language-Codec. Language-Codec is optimized using the AdamW optimizer with an initial learning rate of 2e-4 and betas set to (0.9, 0.999). The learning rate was decayed based on a cosine schedule. MobileSpeech \cite{ji2024mobilespeech} was trained for 12 epochs on 8 NVIDIA A100 40G GPUs, with each batch accommodating 3500 frames of the discrete codecs. We optimized the models using the AdamW optimizer with parameters $\beta _{1}$ = 0.9 and $\beta_{2}$ = 0.95. The learning rate was warmed up for the first 5k updates, reaching a peak of $5\times 10^{-4}$, and then linearly decayed. The AR model and NAR model in VALL-E \cite{valle} are trained using 4 NVIDIA A100 40GB GPUs with a batch size of 6k acoustic tokens per GPU. We also optimize the models with the AdamW optimizer, warm up the learning rate for the first 5k updates to a peak of $5\times 10^{-4}$, and then linear decay it.

\section{Human evaluation}
\label{appendix human}
We conduct the MOS (mean opinion score) evaluation on the Librispeech test set to measure the audio naturalness via crowdsourcing in zero-shot TTS experiments. We keep the text content and prompt speech consistent among different models to exclude other interference factors. We randomly choose 50 samples from the test set for the subjective evaluation and each audio is listened to by at least 10 testers. We analyze the MOS in three aspects: MOS-Q (Quality: clarity, high-frequency, and original timbre reconstruction), MOS-P (Prosody: naturalness of pitch, energy, and duration), and MOS-S (Speaker similarity). 
We require crowdsource evaluators to focus solely on the MOS results for a specific dimension, disregarding the influence of other dimensions. 

\section{More Reconstruction Evaluation}
\label{appendix generalization}
We evaluate the performance of different codec models in noisy environments using the LibriTTS Test-Other dataset. The experimental results are presented in Table \ref{table_testother}. We noticed that all comparative models maintain similar conclusions and trends between the Test-Clean (clean dataset) and Test-Other (noisy dataset) conditions. Moreover, the Language-Codec model demonstrates good reconstruction quality even in noisy environments.

We validated the generalization performance of the codec model on a total of 13,100 audio samples from the LJSpeech dataset. The audio in the LJSpeech dataset has a sampling rate of 22,050 Hz, which we resampled to 24,000 Hz during the input stage of the inference process. Since the codec model was trained on tens of thousands of hours of speech data, it possesses a stronger generalization capability. From Table \ref{table3}, we observed the following findings: 1) Most codec models demonstrated impressive generalization performance, with SpeechTokenizer exhibiting slightly lower generalization performance, likely due to training data limitations. 2) In tests involving different sampling rates and out-of-domain samples, Language-Codec outperformed the current SOTA baseline models significantly across various objective metrics such as UTMOS, PESQ, STOI, V/UV F1, and SPK, for both the four-channel and eight-channel configurations. In general, due to its data-driven nature, Language-Codec demonstrates excellent generalization performance.



\section{Ablation experiments about discriminator}
\label{appendix discriminator}

We evaluated the impact of the ablation experiments on the multi-scale discriminator on the LibriTTS testclean dataset. The experimental results are presented in Table \ref{table8}. We observed that the multi-scale discriminator enhances the reconstruction performance of the Codec model to a certain extent. Specifically, the multi-scale discriminator shows significant improvements in the PESQ metric.

\section{Ablation experiments about decoder}

\label{appendix decoder}

We conducted detailed ablation experiments on the Decoder module within the Language-Codec framework. Specifically, we initially replaced the Decoder with the upsampling module from the Encodec model. This framework was denoted as 'Language-Codec w/o decoder'. Additionally, we examined the impact of the attention module on Codec reconstruction fidelity. This framework was denoted as 'Language-Codec w/o attention block'. The experimental results, as shown in Table \ref{table9}, highlight the significant influence of the upsampling structure dependent on Fourier transforms and the attention module on Codec reconstruction efficacy.

\section{Experiments on downstream generative models with four quantizers}
\label{appendix four quantizer}

Due to time constraints, we evaluated the generation performance of a four-layer quantizer on the MobileSpeech model. The experimental results, presented in Table \ref{table11}, demonstrate that even with just four codebooks, high-quality audio can be generated.




\section{Visualize modules in Decoder}
\label{appendix visual}

As shown in Figure \ref{figurejsp_appendix}(a), the Attention Block is composed of stacked Conv1d ResBlocks and Self-Attention modules. 

Specifically, as shown in Figure \ref{figurejsp_appendix}(b), The Conv1d ResBlock processes the features by first applying normalization and sigmoid activation, followed by a Conv1d layer with the kernel size of 3, stride of 1, and padding of 1. After another round of normalization and sigmoid activation, the features undergo dropout before being passed through another Conv1d layer with the same configuration. Finally, residual connections are applied to produce the output.

The self-attentive module, as shown in Figure \ref{figurejsp_appendix}(c), utilizes three Conv1d layers with kernel size of 1, stride of  1, and padding of 0 to extract Q, K, and V, respectively. After that execute the attention mechanism and finally go through the Conv1d layer output with kernal size of 1, Stride of 1 and padding of 0.

As shown in Figure \ref{figurejsp_appendix}(d), ConvNeXt consists of the stack of N ConvNeXtBlocks, each ConvNeXtBlock processes the input features using the Conv1d layer with Kernal size of 7, Stride of 3, and padding of 1. After Normalize and Linear layer, the output features are processed by GLUE activation and the Linear layer, finally using the residual connection for the input of the next ConvNeXtBlock.

\end{document}